\documentclass[12pt]{article}
\usepackage{amsmath,amssymb}
\usepackage{graphicx}
\newcommand{\eqdef}{\stackrel{\text{def}}{=}}

\newcommand{\n}{\nonumber \\}
\newcommand{\bm}{\boldsymbol}
\newcommand{\ignore}[1]{}

\numberwithin{equation}{section}
\newcommand{\Romannumeral}[1]{\uppercase\expandafter{\romannumeral#1}}
\newcommand{\I}{\text{\Romannumeral{1}}}

\newtheorem{theo}{\bf Theorem}[section]

\newtheorem{prop}[theo]{\bf Proposition}
\newtheorem{defi}[theo]{\bf Definition}
\newcommand{\ma}{\hspace{0pt}}

\newcommand{\cX}{\mathcal{X}}
\allowdisplaybreaks[4]

\begin{document}

\baselineskip=20pt
\newcommand{\preprint}{
\vspace*{-20mm}\begin{flushleft}\end{flushleft}
}
\newcommand{\Title}[1]{{\baselineskip=26pt
  \begin{center} \Large \bf #1 \\ \ \\ \end{center}}}
\newcommand{\Author}{\begin{center}
  \large \bf 
  Ryu Sasaki${}$ \end{center}}
\newcommand{\Address}{\begin{center}
     Department of Physics and Astronomy, Tokyo University of Science,
     Noda 278-8510, Japan
        \end{center}}
\newcommand{\Accepted}[1]{\begin{center}
  {\large \sf #1}\\ \vspace{1mm}{\small \sf Accepted for Publication}
  \end{center}}

\preprint
\thispagestyle{empty}

\Title{Lattice fermions  with solvable  wide range interactions}

\Author

\Address
\vspace{1cm}

\begin{abstract}
Exactly solvable (spinless) lattice  fermions with wide range interactions
are constructed explicitly based on {\em exactly solvable  stationary and reversible Markov chains}
$\mathcal{K}^R$  reported  a few years earlier by Odake and myself.
The reversibility of $\mathcal{K}^R$  with the stationary distribution $\pi$
 leads to a positive classical Hamiltonian $\mathcal{H}^R$. 
 The exact solvability of $\mathcal{H}^R$ warrants that of  
 a spinless lattice fermion $c_x$, $c_x^\dagger$,
 $\mathcal{H}^R_f=\sum_{x,y\in\mathcal{X}}c_x^\dagger\mathcal{H}^R(x,y) c_y$ 
 based on the principle advocated recently by myself.
 The reversible Markov chains $\mathcal{K}^R$ are constructed by  convolutions of the
 orthogonality measures of 
the discrete orthogonal polynomials of Askey scheme.
Several explicit examples of the fermion systems with wide range interactions are presented.
\end{abstract}

%
%
\section{Introduction}
\label{sec:intro}
Here I report a simple construction of exactly solvable  
(spinless) fermions with wide range interactions on a
one dimensional integer lattice. 
Compared to  fermion  systems  with the nearest neighbour interactions,
exactly solvable and wide range interactions are relatively hard to fathom.
The goal is achieved, following the general principle advocated in  \cite{solvfermi}, 
by rewriting the stationary and reversible (detail balanced) Markov chain matrix $\mathcal{K}^R(x,y)$
into a positive classical Hamiltonian $\mathcal{H}^R(x,y)$ by a similarity transformation in terms of
the square root of the stationary distribution $\pi(x)$.
The construction of many exactly solvable, stationary 
and reversible Markov chain matrix $\mathcal{K}^R(x,y)$
on one dimensional  integer lattices
has been reported  a few years earlier by Odake and myself \cite{os39}.
This reference will be cited as \I\, in this paper.
Various convolutions of the orthogonality measures of the discrete orthogonal polynomials
of Askey scheme \cite{askey, kls, ismail, os12}
provide the desired forms of the reversible Markov chain matrix $\mathcal{K}^R(x,y)$.
The corresponding eigen polynomials are  the Krawtchouk (K), Hahn (H),  $q$-Hahn ($q$H),
Meixner (M) and  Charlier (C).

This paper is organised as follows. In section two, the general setting of the stationary and 
reversible Markov chains is recapitulated
with the simple derivation  of the classical  Hamiltonian  $\mathcal{H}^R$.
In \S\ref{sec:convlist} three types of convolutions for constructing 
the reversible Markov chains $\mathcal{K}^R$
are displayed. In \S\ref{sec:Hf} the exactly solvable fermion Hamiltonian $\mathcal{H}^R_f$
with wide range interactions  is trivially constructed from the
classical Hamiltonian $\mathcal{H}^R$.
The main results, several explicit forms of the classical Hamiltonians $\mathcal{H}^R$ with the eigensystems,
are displayed in section four. They are grouped for each orthogonal polynomial.
Those for the finite polynomials are shown first, with three forms of $\mathcal{K}^R$ corresponding to 
the convolution types.  
The corresponding Hamiltonians belonging to the infinite polynomials are shown after them.
They are obtained by certain limit procedures from those of the finite polynomials.

\section{General setting; stationary and reversible Markov chains $\mathcal{K}^R$}
\label{sec:set}

Let us start with a brief recapitulation of the general setting 
of classical stationary Markov chains $\mathcal{K}$
on a one dimensional integer lattice $\mathcal{X}$, finite or semi-infinite,
\begin{equation}
x,y,z\in \mathcal{X}=\{0,1,\ldots,N\}, \quad N\in\mathbb{N},
\quad x,y,z\in \mathcal{X}=\mathbb{Z}_{\ge0},
\label{lattice}
\end{equation}
and points on $\mathcal{X}$ are denoted by $x,y,z$ for analytical treatments.
I use the convention that the transition probability matrix per 
unit time interval $\mathcal{K}(x,y)$ on $\mathcal{X}$ means
the transition from an initial point $y$ to a final point $x$ 
with $ \mathcal{K}(x,y)>0$ and it satisfies the conservation of
the probability
 \begin{equation}
 \sum_{x\in{X}}\mathcal{K}(x,y)=1.
  \label{consP}
\end{equation}
The positive $\mathcal{K}$ means that all the points on $\mathcal{X}$ 
are connected with each other by $\mathcal{K}$ 
and this translates into  wide range interactions.
\begin{defi}
\label{def:rev}
Markov chain $\mathcal{K}^R$ is reversible {\rm (}detail balanced\/{\rm)} 
if it has a reversible  distribution $\pi$ satisfying
\begin{align}
&\mathcal{K}^R(x,y)\pi(y)=\mathcal{K}^R(y,x)\pi(x),\quad x,y\in\mathcal{X};\quad 
\pi(x)>0,\quad \sum_{x\in\mathcal{X}}\pi(x)=1.
\label{revdef}
\end{align}
\end{defi}
Taking $y$ summation of the reversibility definition \eqref{revdef} 
\begin{align}
\sum_{y\in\mathcal{X}}\mathcal{K}^R(x,y)\pi(y)
=\pi(x)\sum_{y\in\mathcal{X}}\mathcal{K}^R(y,x)=\pi(x).
\label{stdis}
\end{align}
leads to the following 
\begin{prop}
\label{eigenpi}
The reversible $\mathcal{K}^R(x,y)$ has
 a Perron-Frobenius eigenvector $\pi(x)$ with the maximal and simple eigenvalue 1   
 and the range of spectrum
\begin{equation*}
  -1<\text{The moduli of the eigenvalues of }\mathcal{K}^R(x,y)\le1.
\end{equation*}
\end{prop}
This is a consequence of the positivity, {\em i.e.}\ Perron-Frobenius theorem,
and the probability conservation \eqref{consP}.
\begin{prop}
\label{prop:HRdef}
A positive Hamiltonian $\mathcal{H}^R$, a real symmetric $|\mathcal{X}|\times|\mathcal{X}|$ matrix,
is obtained by dividing the definition of the reversible Markov chain matrix $\mathcal{K}^R$ 
\eqref{revdef} by $\sqrt{\pi(x)}\sqrt{\pi(y)}$,
\begin{equation}
\mathcal{H}^R(x,y)\eqdef \frac1{\sqrt{\pi(x)}}\,\mathcal{K}^R(x,y)\,\sqrt{\pi(y)}
=\frac1{\sqrt{\pi(y)}}\,\mathcal{K}^R(y,x)\,\sqrt{\pi(x)}=\mathcal{H}^R(y,x),\quad x,y\in\mathcal{X}.
\label{HRdef}
\end{equation}
It has the Perron-Frobenius eigenvector $\sqrt{\pi(x)}$,
\begin{equation}
\sum_{y\in\mathcal{X}}\mathcal{H}^R(x,y)\sqrt{\pi(y)}
=\frac1{\sqrt{\pi(x)}}\sum_{y\in\mathcal{X}}\mathcal{K}^R(x,y)\pi(y)
=\sqrt{\pi(x)}\sum_{y\in\mathcal{X}}\mathcal{K}^R(y,x)=\sqrt{\pi(x)},
\end{equation}
and  its eigenvalues are all real,
\begin{equation}
  -1<\text{The  eigenvalues of }\mathcal{H}^R(x,y)\le1.
  \label{eigreal}
\end{equation}
\end{prop}

\section{Construction  of  fermion Hamiltonian $\mathcal{H}^R_f$ with wide range interactions}
\label{sec:constH}

Here I explain the construction method of reversible and {\em finite} Markov chains  $\mathcal{K}^R$
adopted in \I \,\cite{os39}. 
This gives a classical Hamiltonian $\mathcal{H}^R$ \eqref{HRdef}
 and the general principle of \cite{solvfermi}
provides the fermion Hamiltonian $\mathcal{H}^R_f$ with wide range interactions.
Those for the {\em infinite} Markov chains are derived by certain limiting
procedures including $N\to\infty$, as will be shown shortly.
The method is based on certain convolutions of the orthogonality measures of discrete orthogonal
polynomials of Askey scheme. 
Normalised orthogonality measure, being positive, can always be a probability distribution.
Let us denote the normalised orthogonality measure with the explicit $N$ dependence by
\begin{equation}
\pi(x,N,\bm{\lambda})>0,\qquad \sum_{x\in\mathcal{X}}\pi(x,N,\bm{\lambda})=1,
\label{genpi}
\end{equation}
in which $\bm{\lambda}$ stands for the set of parameters.

\subsection{Three types of convolutions}
\label{sec:convlist}

Here I present three types of convolutions among five reported in I \cite{os39} for simplicity and clarity;
\begin{align}
  \text{(\romannumeral1)}:&\ \ \mathcal{K}^R(x,y;\bm{\lambda}_1,\bm{\lambda}_2)
  \eqdef\sum_{z=0}^{\min(x,y)}
  \!\pi(x-z,N-z,\bm{\lambda}_2)\pi(z,y,\bm{\lambda}_1),
  \label{conv11}\\
  \text{(\romannumeral2)}:&\ \ \mathcal{K}^R(x,y;\bm{\lambda}_1,\bm{\lambda}_2)
  \eqdef\sum_{z=\max(0,x+y-N)}^{\min(x,y)}
  \!\!\!\!\!\!\!\!\!\pi(x-z,N-y,\bm{\lambda}_2)\pi(z,y,\bm{\lambda}_1),
  \label{conv2}\\
  \text{(\romannumeral3)}:&\ \ \mathcal{K}^R(x,y;\bm{\lambda}_1,\bm{\lambda}_2)
  \eqdef\sum_{z=\max(x,y)}^N
  \!\!\!\!\pi(x,z,\bm{\lambda}_2)\pi(z-y,N-y,\bm{\lambda}_1).
  \label{conv3}
\end{align}
It is easy to verify the positivity and  the conservation of the probability \eqref{consP} 
for each convolution.
The strategy is to find a good set of parameters $\bm{\lambda}_i$, 
$i=1,2,3$, such that the reversibility condition
\begin{equation}
\mathcal{K}^R(x,y;\bm{\lambda}_1,\bm{\lambda}_2)\pi(y,N,\bm{\lambda}_3)=
\mathcal{K}^R(y,x;\bm{\lambda}_1,\bm{\lambda}_2)\pi(x,N,\bm{\lambda}_3),
\label{revdef2}
\end{equation}
is satisfied. Obviously $\bm{\lambda}_3$ is a function of the parameters 
$\bm{\lambda}_i$, $i=1,2$ in $\mathcal{K}^R$.

The main result of \I\,\cite{os39} is the following 
\begin{theo}
\label{theo:39}
{\rm (}Odake-Sasaki\/{\rm )}\\
The finite discrete orthogonal polynomials 
$\{\check{P}_n(x,\bm{\lambda}_3)\}$, whose orthogonality measure
$\pi(x,N,\bm{\lambda}_3)$ provides the reversible distribution of 
$\mathcal{K}^R(x,y;\bm{\lambda}_1,\bm{\lambda}_2)$, constitute the left eigenvectors of
$\mathcal{K}^R(x,y;\bm{\lambda}_1,\bm{\lambda}_2)$,
\begin{align} 
\sum_{x\in\mathcal{X}}\mathcal{K}^R(x,y;\bm{\lambda}_1,\bm{\lambda}_2)
\check{P}_n(x,\bm{\lambda}_3)=\kappa(n)\check{P}_n(y,\bm{\lambda}_3),
\  -1<\kappa(n)\le1,\  x,n\in\mathcal{X}, \ \kappa(0)=1.
\label{leig}
\end{align}
The right eigenvectors are $\{\pi(x,N,\bm{\lambda}_3)\check{P}_n(x,\bm{\lambda}_3)\}$,
\begin{align} 
\sum_{y\in\mathcal{X}}\mathcal{K}^R(x,y;\bm{\lambda}_1,\bm{\lambda}_2)
\pi(y,N,\bm{\lambda}_3)\check{P}_n(y,\bm{\lambda}_3)
&=
\pi(x,N,\bm{\lambda}_3)\sum_{y\in\mathcal{X}}
\mathcal{K}^R(y,x;\bm{\lambda}_1,\bm{\lambda}_2)\check{P}_n(y,\bm{\lambda}_3)\n
&=\kappa(n)\pi(x,N,\bm{\lambda}_3)\check{P}_n(x,\bm{\lambda}_3),\qquad x,n\in\mathcal{X}.
\label{reig}
\end{align}
The Hamiltonian $\mathcal{H}^R$ has the eigenvectors 
$\{\sqrt{\pi(x,N,\bm{\lambda}_3)}\,\check{P}_n(x,\bm{\lambda}_3)\}$,
\begin{align} 
&\sum_{y\in\mathcal{X}}\mathcal{H}^R(x,y;\bm{\lambda}_1,\bm{\lambda}_2)
\sqrt{\pi(y,N,\bm{\lambda}_3)}\,\check{P}_n(y,\bm{\lambda}_3)\n
&=\frac1{\sqrt{\pi(x,N,\bm{\lambda}_3)}}\sum_{y\in\mathcal{x}}
\mathcal{K}^R(x,y;\bm{\lambda}_1,\bm{\lambda}_2)
\pi(y,N,\bm{\lambda}_3)\,\check{P}_n(y,\bm{\lambda}_3)\n
&=\kappa(n)\sqrt{\pi(x,N,\bm{\lambda}_3)}\,\check{P}_n(y,\bm{\lambda}_3).
\label{Heig}
\end{align}
\end{theo}
The normalisation constant  of the polynomial $\check{P}_n(x)$ 
is determined by the universal normalisation condition
\begin{align} 
\check{P}_n(0,\bm{\lambda}_3)&=1,\qquad \quad \forall n\in\mathcal{X},
\label{univnorm}\\
\sum_{x\in\mathcal{X}}\pi(x,N,\bm{\lambda}_3)
\check{P}_m(x,\bm{\lambda}_3)\check{P}_n(x,\bm{\lambda}_3)&=\frac{\delta_{m,n}}{d_n^2},
\quad d_n>0, 
\quad m,n\in\mathcal{X}.
\end{align}
Of course, the constant $d_n$ also depends on $\bm{\lambda}_3$ but its dependence is suppressed for simplicity
of presentation. 
It should be noted that, because of the context, the present definition of
$d_n^2$ is slightly different from  previous one
\cite{solvfermi,os12}.
The orthonormal eigenvectors of the classical Hamiltonian $\mathcal{H}^R$ are $\hat{\phi}_n(x)$,
\begin{align} 
&\mathcal{H}^R\hat{\phi}_n=\kappa(n)\hat{\phi}_n\ \Leftrightarrow
\sum_{y\in\mathcal{X}}\mathcal{H}^R(x,y;\bm{\lambda}_1,\bm{\lambda}_2)
\hat{\phi}_n(y)=\kappa(n)\hat{\phi}_n(x),\n
&\qquad \hat{\phi}_n(x)\eqdef d_n\sqrt{\pi(x,N,\bm{\lambda}_3)}
\,\check{P}_n(x,\bm{\lambda}_3)\in\mathbb{R},
\label{phindef}\\
&\sum_{x\in\mathcal{X}}\hat{\phi}_m(x)\hat{\phi}_n(x)=\delta_{m,n},\quad
\sum_{n\in\mathcal{X}}\hat{\phi}_n(x)\hat{\phi}_n(y)=\delta_{x,y}.
\label{ornor}
\end{align}

\subsection{Fermion Hamiltonian $\mathcal{H}^R_f$}
\label{sec:Hf}

The fermion Hamiltonian $\mathcal{H}^R_f$ with wide range interactions is defined from the 
classical Hamiltonian $\mathcal{H}^R$ as 
a bi-linear form of the lattice fermions $\{c_x\}$, $\{c_x^\dagger\}$ on $\mathcal{X}$,  obeying the
canonical anti-commutation relations,
\begin{align}
 \{c_x^\dagger,c_y\}&=\delta_{x,y},\quad \{c_x^\dagger,c_y^\dagger\}=0=\{c_x,c_y\},
\quad x,y\in\mathcal{X},
\label{comrel}\\
\mathcal{H}^R_f&\eqdef \sum_{x,y\in\mathcal{X}}c_x^\dagger\mathcal{H}^R(x,y)c_y,
\label{Hf0}
\end{align}
in which the parameter dependence is suppressed for simplicity of presentation.
\begin{theo}
\label{theo:main}
The Hamiltonian  $\mathcal{H}^R_f$ is diagonalised by the introduction of the 
the momentum space fermion operators 
$\{\hat{c}_n\}$, $\{\hat{c}_n^\dagger\}$, $n\in\mathcal{X}$,
\begin{align}
\hat{c}_n&\eqdef\sum_{x\in\mathcal{X}}\hat{\phi}_n(x)c_x,\ \
\hat{c}_n^\dagger=\sum_{x\in\mathcal{X}}\hat{\phi}_n(x)c_x^\dagger\ \Leftrightarrow \
{c}_x=\sum_{n\in\mathcal{X}}\hat{\phi}_n(x)\hat{c}_n,\ \
{c}_x^\dagger=\sum_{m\in\mathcal{X}}\hat{\phi}_m(x)\hat{c}_m^\dagger,
\label{momrep}\\
&\hspace{2cm} \Longrightarrow \ \{\hat{c}_m^\dagger,\hat{c}_n\}
=\delta_{m\,n},\ \{\hat{c}_m^\dagger,\hat{c}_n^\dagger\}=0=\{\hat{c}_m,\hat{c}_n\},
\label{chatcom0}\\[-2pt]
&\hspace{6cm} \Downarrow \n[-2pt]
 &\mathcal{H}^R_f=\sum_{m,n,x,y\in\mathcal{X}}
\hat{\phi}_m(x)\mathcal{H}^R(x,y)\hat{\phi}_n(y)\hat{c}_m^\dagger\hat{c}_n
=\sum_{m,n,x\in\mathcal{X}}\kappa(n)\hat{\phi}_m(x)\hat{\phi}_n(x)\hat{c}_m^\dagger\hat{c}_n\n
&\phantom{\mathcal{H}^R_f}=\sum_{n\in\mathcal{X}}\kappa(n)\hat{c}_n^\dagger\hat{c}_n,
\label{Hjdiag}\\
&\hspace{2cm} \Longrightarrow \ [\mathcal{H}^R_{f},\hat{c}_n^\dagger]=\kappa(n)\hat{c}_n^\dagger,
\qquad [\mathcal{H}^R_{f},\hat{c}_n]=-\kappa(n)\hat{c}_n.
\label{Hfcom10}
\end{align}
\end{theo}

\section{Explicit forms of the classical Hamiltonians  $\mathcal{H}^R$}
\label{sec:expH}

Here I present the explicit forms of the reversible Markov chain matrices $\mathcal{K}^R(x,y)$
belonging to certain  subset of the discrete orthogonal polynomials of 
Askey scheme \cite{askey,kls,ismail,os12}.
They are all reported in \I \,\cite{os39} and reproduced here for self-containedness.
The polynomials are the Krawtchouk (K), Charlier (C), Hahn (H), Meixner (M)  and $q$-Hahn ($q$H).
For each polynomial, after listing the basic data,  at most three types of  reversible
Markov chain matrices $\mathcal{K}^R(x,y)$ are displayed.
The classical Hamiltonian $\mathcal{H}^R$ \eqref{HRdef} 
and the fermion Hamiltonian $\mathcal{H}^R_f$ \eqref{Hf0} are obtained straightforwardly.
They could be used to calculate many interesting quantities of the  fermions with wide range interactions, 
{\em e.g.} entanglement entropy, etc
\cite{gvz,ortho,Kra,finkel,latorre, qcmarkov}.

\subsection{Krawtchouk}
\label{sec:Kra}

The polynomial depends on one positive parameter $\bm{\lambda}=p$ ($0<p<1$),
\begin{align}
  &\pi(x,N,p)=\binom{N}{x}p^x(1-p)^{N-x},\quad
  \binom{N}{x}=\frac{N!}{x!\,(N-x)!}\,,\quad
  d_n^2=\binom{N}{n}\Bigl(\frac{p}{1-p}\Bigr)^n,\\[2pt]
  &\check{P}_n(x,p)=P_n(x,p)
  ={}_2F_1\Bigl(\genfrac{}{}{0pt}{}{-n,\,-x}{-N}\Bigm|p^{-1}\Bigr),\quad
  P_n(x,p)=P_x(n,p),\quad     \text{(self-dual)}.
  \label{Kp}
\end{align}
\subsubsection{Type (i) convolution} 
\label{sec:Kra1}
This convolution has been applied to (K) and (H) in many papers
\cite{hoa-rah83}--\cite{albert} in connection with ``cumulative Bernoulli trials.''
By taking $\bm{\lambda}_1=a$, $\bm{\lambda}_2=b$  and
$\bm{\lambda}_3=p\eqdef\frac{b}{1-a+ab}$,
the matrix $\mathcal{K}^R(x,y)$ is
\begin{equation}
\mathcal{K}^R(x,y)=\!\sum_{z=0}^{\min(x,y)}\pi(x-z,N-z,b)\pi(z,y,a),
\quad 0<a,b,p<1,
  \label{KK1}
\end{equation}
satisfying
\begin{align}
  &\sum_{y\in\mathcal{X}}\mathcal{K}^R(x,y)\pi(y,N,p)\check{P}_n(y,p)
  =\kappa(n)\pi(x,N,p)\check{P}_n(x,p), \n
 &\hspace{3cm} \kappa(n)=a^n(1-b)^n
  ={}_1F_0\Bigl(\genfrac{}{}{0pt}{}{-n}{-}\Bigm|bp^{-1}\Bigr),\quad n\in\cX.
  \label{Pk1}\\
&\Rightarrow \mathcal{H}^R(x,y)=\frac1{\sqrt{\pi(x,N,p)}}
\!\sum_{z=0}^{\min(x,y)}\!\!\pi(x-z,N-z,b)\,\pi(z,y,a)
\sqrt{\pi(y,N,p)},
\label{K1H}\\[2pt]
&\qquad \hat{\phi}_n(x)
=d_n\sqrt{\pi(x,N,p)}\,{}_2F_1\Bigl(\genfrac{}{}{0pt}{}{-n,\,-x}{-N}\Bigm|p^{-1}\Bigr),
\quad d_n^2={\scriptsize 
\bigl(
\begin{array}{l}
 N  \\
n
\end{array}
\bigr)
}
(\tfrac{p}{1-p})^n, \quad p=\tfrac{b}{1-a+ab}.
\label{K1phn}
\end{align}

\subsubsection{Type (ii) convolution}
\label{sec:Kra2}
By taking $\bm{\lambda}_1=a$, $\bm{\lambda}_2=b$ and
$\bm{\lambda}_3=p\eqdef\frac{b}{1-a+b}$,
the matrix $\mathcal{K}^R(x,y)$ is
\begin{equation}
\mathcal{K}^R(x,y)=\sum_{z=\max(0,x+y-N)}^{\min(x,y)}\pi(x-z,N-y,b)\pi(z,y,a),
\quad 0<a,b,p<1,
  \label{KK2}
\end{equation}
satisfying
\begin{align}
 & \sum_{y\in\mathcal{X}}\mathcal{K}^R(x,y)\pi(y,N,p)\check{P}_n(y,p)
  =\kappa(n)\pi(x,N,p)\check{P}_n(x,p), \n
&\hspace{3cm}  \kappa(n)=(a-b)^n
  ={}_1F_0\Bigl(\genfrac{}{}{0pt}{}{-n}{-}\Bigm|bp^{-1}\Bigr).\quad n\in\cX,
  \label{Pk2}\\
 &\Rightarrow \mathcal{H}^R(x,y)=\frac1{\sqrt{\pi(x,N,p)}}
 \!\sum_{z=\max(0,x+y-N)}^{\min(x,y)}\pi(x-z,N-y,b)\pi(z,y,a)\sqrt{\pi(y,N,p)},
\label{K2H}\\[2pt]
&\qquad \hat{\phi}_n(x)=d_n\sqrt{\pi(x,N,p)}\,
{}_2F_1\Bigl(\genfrac{}{}{0pt}{}{-n,\,-x}{-N}\Bigm|p^{-1}\Bigr),
\quad d_n^2={\scriptsize 
\bigl(
\begin{array}{l}
 N  \\
n
\end{array}
\bigr)
}
(\tfrac{p}{1-p})^n, \quad p=\tfrac{b}{1-a+b}.
\label{K2phn}
\end{align}
It is interesting to note that odd eigenvalues are all negative if $0<a<b<1$.
\subsubsection{Type (iii) convolution} 
\label{sec:Kra3}
By taking $\bm{\lambda}_1=a$, $\bm{\lambda}_2=b$ and
$\bm{\lambda}_3=p\eqdef\frac{ab}{1-b+ab}$,
the matrix $\mathcal{K}^R(x,y)$ is
\begin{align}
\mathcal{K}^R(x,y)&=\sum_{z=\max(x,y)}^N\!\!\pi(x,z,b)\pi(z-y,N-y,a)\quad
0<a,b,p<1.
  \label{KK3}
\end{align}
satisfying
\begin{align}
  &\sum_{y\in\mathcal{X}}\mathcal{K}^R(x,y)\pi(y,N,p)\check{P}_n(y,p)
  =\kappa(n)\pi(x,N,p)\check{P}_n(x,p), \n
 &\hspace{3cm} \kappa(n)=(1-a)^nb^n
  = {}_1F_0\Bigl(\genfrac{}{}{0pt}{}{-n}{-}\Bigm|abp^{-1}\Bigr),\quad n\in\cX,
  \label{Pk3}\\
 &\Rightarrow \mathcal{H}^R(x,y)=\frac1{\sqrt{\pi(x,N,p)}}\!
 \sum_{z=\max(x,y)}^N\pi(x,z,b)\pi(z-y,N-y,a)\sqrt{\pi(y,N,p)},
\label{K3H}\\[2pt]
&\qquad \hat{\phi}_n(x)=d_n\sqrt{\pi(x,N,p)}\,{}_2F_1\Bigl(\genfrac{}{}{0pt}{}{-n,\,-x}{-N}\Bigm|p^{-1}\Bigr),
\quad d_n^2={\scriptsize 
\bigl(
\begin{array}{l}
 N  \\
n
\end{array}
\bigr)
}
(\tfrac{p}{1-p})^n, \quad p=\tfrac{ab}{1-a+ab}.
\label{K3phn}  
\end{align}
\subsection{Charlier}
\label{sec:Cha}

This polynomial is defined on a semi-infinite integer lattice
$\cX=\mathbb{Z}_{\ge0}$ depending on one positive parameter $\bm{\lambda}=a$ ($a>0$),
\begin{align}
  &\pi(x,a)=\frac{a^xe^{-a}}{x!},\quad
  d_n^2=\frac{a^n}{n!},
  \label{Cpid}\\
  &\check{P}_n(x,a)=P_n(x,a)
  ={}_2F_0\Bigl(\genfrac{}{}{0pt}{}{-n,\,-x}{-}\Bigm|-a^{-1}\Bigr),\quad
  P_n(x,a)=P_x(n,a),\quad \text{(self-dual)}.
  \label{Cp}
\end{align}
By the replacement $p\to pN^{-1}$ and the limit $N\to\infty$, the
Krawtchouk (K) goes to Charlier (C) \cite{kls},
\begin{equation*}
  \check{P}_{\text{K}\,n}(x,p)\to\check{P}_{\text{C}\,n}(x,p),\quad
  \pi_{\text{K}}(x,N,p)\to\pi_{\text{C}}(x,p),\quad
  d_{\text{K}\,n}^2\to d_{\text{C}\,n}^2.
\end{equation*}
\subsubsection{Type (i) convolution}
\label{sec:Cha1}
This is achieved by $b\to bN^{-1}$, $N\to\infty$ in  
$\mathcal{K}^R$ of Krawtchouk type (i) convplution \eqref{KK1},
\begin{align}
  &\check{P}_n(x,p)\to\check{P}_{\text{C}\,n}(x,p'),\quad
  p'\eqdef\frac{b}{1-a}, \qquad \qquad 0<a<1,\n
  &\pi(x,N,p)\to\pi_{\text{C}}(x,p'),\quad
  \kappa(n)\to\kappa_{\text{C}}(n)=a^n,\n
  &\mathcal{K}^R(x,y)\to\mathcal{K}^R_{\text{C}}(x,y)
  =\sum_{z=0}^{\min(x,y)}\pi_{\text{C}}(x-z,b)\pi_{\text K}(z,y,a),
  \label{KCconv1}\\
& \hspace{-1cm} \Longrightarrow\mathcal{H}^R(x,y)=\frac1{\sqrt{\pi_\text{C}(x,p')}}
\sum_{z=0}^{\min(x,y)}\pi_{\text{C}}(x-z,b)\pi_{\text K}(z,y,a)\sqrt{\pi_\text{C}(y,p')},
\label{C1H}\\
&\hat{\phi}_n(x)=d_n\pi_\text{C}(x,p')\,
{}_2F_0\Bigl(\genfrac{}{}{0pt}{}{-n,\,-x}{-}\Bigm|-{p'}^{-1}\Bigr),\quad d_n^2={p'}^n/n!.
\label{C1phn}
\end{align}
The infinite limit of Krawtchouk $\mathcal{K}^R(x,y)$ of Type (ii) convolution gives the same result as this one.
\subsubsection{Type (iii) convolution}
\label{sec:Cha3}

This is achieved by $a\to aN^{-1}$, $N\to\infty$  in  
$\mathcal{K}^R$ of Krawtchouk type (iii) convplution \eqref{KK3},
\begin{align}
  &\check{P}_n(x,p)\to P_{\text{C}\,n}(x,p'),\quad
  p'\eqdef\frac{ab}{1-b}, \qquad \qquad 0<b<1,\n
  &\pi(x,N,p)\to\pi_{\text{C}}(x,p'),\quad
  \kappa(n)\to\kappa_{\text{C}}(n)=b^n={}_1F_0\Bigl(\genfrac{}{}{0pt}{}{-n}{-}\Bigm|abp^{\prime\,-1}\Bigr),\n
  &\mathcal{K}^R(x,y)\to \mathcal{K}^R_{\text{C}}(x,y)
  =\sum_{z=\max(x,y)}^{\infty}\pi_{\text K}(x,z,b)\pi_{\text{C}}(z-y,a),
  \label{KCconv3}\\
& \hspace{-1cm} \Longrightarrow\mathcal{H}^R(x,y)=\frac1{\sqrt{\pi_\text{C}(x,p')}}
\sum_{z=\max(x,y)}^{\infty}\pi_{\text K}(x,z,b)\pi_{\text{C}}(z-y,a)\sqrt{\pi_\text{C}(y,p')},
\label{C3H}\\
&\hat{\phi}_n(x)=d_n\pi_\text{C}(x,p')\,
{}_2F_0\Bigl(\genfrac{}{}{0pt}{}{-n,\,-x}{-}\Bigm|-{p'}^{-1}\Bigr),\quad d_n^2={p'}^n/n!.
\label{C3phn}
\end{align}
\subsection{Hahn}
\label{sec:Hah}
The polynomial depends on two positive parameters $\bm{\lambda}=(a,b)$
($a,b>0$),
\begin{align}
  &\pi(x,N,a,b)=\binom{N}{x}\frac{(a)_x\,(b)_{N-x}}{(a+b)_N},\quad
  d_n^2=\binom{N}{n}\frac{(a)_n\,(2n+a+b-1)(a+b)_N}{(b)_n\,(n+a+b-1)_{N+1}},\\[4pt]
  &\check{P}_n(x,a,b)=P_n(x,a,b)
  ={}_3F_2\Bigl(\genfrac{}{}{0pt}{}{-n,\,n+a+b-1,\,-x}{a,\,-N}\Bigm|1\Bigr).
  \label{Hp}
\end{align}
\subsubsection{Type (i) convolution}
\label{sec:Hah1}

For $\bm{\lambda}_1=(a,b)$,  $\bm{\lambda}_2=(b,c)$ and  $\bm{\lambda}_3=(a+b,c)$, the matrix
$\mathcal{K}^R(x,y)$ is
\begin{equation}
 \mathcal{K}^R(x,y)=\!\sum_{z=0}^{\min(x,y)}\pi(x-z,N-z,b,c)\pi(z,y,a,b),
 \qquad 0<a,b,c,
  \label{h12K}
\end{equation}
 satisfying
\begin{align}
 & \sum_{y\in\cX} \mathcal{K}^R(x,y)\pi(y,N,a+b,c)\check{P}_n(y,a+b,c)
  =\kappa(n)\pi(x,N,a+b,c)\check{P}_n(x,a+b,c),\n
&\qquad    \kappa(n)=\frac{(a)_n(c)_n}{(a+b)_n(b+c)_n}
={}_3F_2\Bigl(
  \genfrac{}{}{0pt}{}{-n,\,n+a+b+c-1,\,b}{a+b,\,b+c}\Bigm|1\Bigr),\quad n\in\cX,
\label{Hah1ka}\\[2pt]
&\hspace{-1cm} \Longrightarrow\mathcal{H}^R(x,y)=\frac1{\sqrt{\pi(x,N,a+b,c)}}\!\!\!\!
\sum_{z=0}^{\min(x,y)}\!\!\!\pi(x-z,N-z,b,c)\pi(z,y,a,b)\sqrt{\pi(y,N,a+b,c)},
\label{HH1}\\[2pt]
&\hat{\phi}_n(x)=d_n\pi(x,N,a+b,c)
\,{}_3F_2\Bigl(\genfrac{}{}{0pt}{}{-n,\,n+a+b+c-1,\,-x}{a+b,\,-N}\Bigm|1\Bigr),
\label{H1phn}\\[2pt]
&\qquad  d_n^2=\binom{N}{n}\frac{(a+b)_n\,(2n+a+b+c-1)(a+b+c)_N}{(c)_n\,(n+a+b+c-1)_{N+1}}.
\end{align}

\subsubsection{Type (ii) convolution}
\label{sec:Hah2}
For $\bm{\lambda}_1=(a,b)$,  $\bm{\lambda}_2=(b,c)$ and $\bm{\lambda}_3=(a+b,b+c)$ the matrix
$\mathcal{K}^R(x,y)$ is
\begin{equation}
 \mathcal{K}^R(x,y)=\!\!\!\sum_{z=\max(0,x+y-N)}^{\min(x,y)}\!\!\!
  \pi(x-z,N-y,b,c)\pi(z,y,a,b), 
  \label{HKt2}
\end{equation}
satisfying 
\begin{align}
 & \sum_{y\in\cX} \mathcal{K}^R(x,y)\pi(y,N,a+b,b+c)\check{P}_n(y,a+b,b+c)\n
  &\qquad \qquad =\kappa(n)\pi(x,N,a+b,b+c)\check{P}_n(x,a+b,b+c),\n
&\qquad    
\kappa(n)=\sum_{k=0}^n\binom{n}{k}(-1)^k\frac{(b)_k(n+a+2b+c-1)_k}{(a+b)_k(b+c)_k}
\label{Hah2ka}\\
&\qquad \phantom{ \kappa(n)}=
{}_3F_2\Bigl(
  \genfrac{}{}{0pt}{}{-n,\,n+a+2b+c-1,\,b}{a+b,\,b+c}\Bigm|1\Bigr),\quad n\in\cX,\\
&\hspace{-1cm} \Longrightarrow\mathcal{H}^R(x,y)=\frac1{\sqrt{\pi(x,N,a+b,b+c)}}\!
\sum_{z=\max(0,x+y-N)}^{\min(x,y)}\!\!\!
  \pi(x-z,N-y,b,c)\pi(z,y,a,b)\n
&\hspace{8cm}\times \sqrt{\pi(y,N,a+b,b+c)},
\label{HH2}\\
&\hat{\phi}_n(x)=d_n\pi(x,N,a+b,b+c)
\,{}_3F_2\Bigl(\genfrac{}{}{0pt}{}{-n,\,n+a+2b+c-1,\,-x}{a+b,\,-N}\Bigm|1\Bigr),
\label{H2phn}\\[2pt]
&\qquad  
d_n^2=\binom{N}{n}\frac{(a+b)_n\,(2n+a+2b+c-1)(a+2b+c)_N}{(b+c)_n\,(n+a+2b+c-1)_{N+1}}.
\end{align}

\subsubsection{Type (iii) convolution}
\label{sec:Hah3}

For $\bm{\lambda}_1=(a,b)$,  
$\bm{\lambda}_2=(c,a)$ and $\bm{\lambda}_3=(c,a+b)$, the matrix
$ \mathcal{K}^R(x,y)$ is
\begin{equation}
 \mathcal{K}^R(x,y)=\!\!\!\sum_{z=\max(x,y)}^N\!\!\!\pi(x,z,c,a)\pi(z-y,N-y,a,b)
  \quad 0<a,b,c,
  \label{HKt3}
\end{equation}
satisfying 
\begin{align}
 & \sum_{y\in\cX} \mathcal{K}^R(x,y)\pi(y,N,c,a+b)\check{P}_n(y,c,a+b)\n
  &\qquad \qquad =\kappa(n)\pi(x,N,c,a+b)\check{P}_n(x,c,a+b),\n[2pt]
&\kappa(n)=\frac{(b)_n(c)_n}{(a+b)_n(a+c)_n}=
{}_3F_2\Bigl(\genfrac{}{}{0pt}{}{-n,\,n+a+b+c-1,\,a}{a+b,\,a+c}\Bigm|1\Bigr),\quad n\in\cX,
\label{Hah3ka}\\
&\hspace{-1cm} \Longrightarrow\mathcal{H}^R(x,y)=\frac1{\sqrt{\pi(x,N,c,a+b)}}\!
\sum_{z=\max(x,y)}^N\!\!\!\pi(x,z,c,a)\pi(z-y,N-y,a,b)
 \sqrt{\pi(y,N,c,a+b)},
\label{HH3}\\
&\hat{\phi}_n(x)=d_n\pi(x,N,c,a+b)
\,{}_3F_2\Bigl(\genfrac{}{}{0pt}{}{-n,\,n+a+b+c-1,\,-x}{c,\,-N}\Bigm|1\Bigr),
\label{H3phn}\\[2pt]
&\qquad  
d_n^2=\binom{N}{n}\frac{(c)_n\,(2n+a+b+c-1)(a+b+c)_N}{(a+b)_n\,(n+a+b+c-1)_{N+1}}.
\end{align}
\subsection{Meixner}
\label{sec:Mei}
This self-dual polynomial is defined on a semi-infinite integer lattice
$\cX=\mathbb{Z}_{\ge0}$ with two positive parameters $\bm{\lambda}=(a,b)$ ($0<a$, $0<b<1$),
\begin{align}
  &\pi(x,a,b)=\frac{(a)_x\,b^x(1-b)^a}{x!},\quad
  d_n^2=\frac{(a)_n\,b^n}{n!},
  \label{Mpi}\\
  &\check{P}_n(x,a,b)=P_n(x,a,b)
  ={}_2F_1\Bigl(\genfrac{}{}{0pt}{}{-n,\,-x}{a}\Bigm|1-b^{-1}\Bigr),\quad
  P_n(x,a,b)=P_x(n,a,b).
  \label{Mp}
\end{align}
By the replacement $b\to N(1-b)b^{-1}$ and the limit $N\to\infty$, the Hahn (H)
goes to Meixner (M),
\begin{equation*}
  \check{P}_{\text{H}\,n}(x,a,b)\to\check{P}_{\text{M}\,n}(x,a,b),\quad
  \pi_{\text{H}}(x,N,a,b)\to\pi_{\text{M}}(x,a,b),\quad 
  d_{\text{H}\,n}^2\to d_{\text{M}\,n}^2.
\end{equation*}
By the replacement $b\to b/(a+b)$ and the limit $a\to\infty$, the Meixner
(M) goes to Charlier (C)
\begin{equation*}
  \check{P}_{\text{M}\,n}(x,a,b)\to\check{P}_{\text{C}\,n}(x,b),\quad
  \pi_{\text{M}}(x,a,b)\to\pi_{\text{C}}(x,b),\quad 
  d_{\text{M}\,n}^2\to d_{\text{C}\,n}^2.
\end{equation*}
\subsubsection{Type (i) convolution}
\label{sec:Mei1}

This is achieved by fixing $a$ and $b$ with $c\to N(1-c)c^{-1}$
($\Rightarrow 0<c<1$), $N\to\infty$ in $\mathcal{K}^R$ of Hahn type (i) convolution \eqref{h12K},
\begin{align}
  &\check{P}_n(x,a+b,c)\to\check{P}_{\text{M}\,n}(x,a+b,c),\n
  &\pi(x,N,a+b,c)\to\pi_{\text{M}}(x,a+b,c),\quad
  \kappa(n)\to\kappa_{\text{M}}(n)=\frac{(a)_n}{(a+b)_n}={}_2F_1\Bigl(
  \genfrac{}{}{0pt}{}{-n,\,b}{a+b}\Bigm|1\Bigr),
\label{HMka}\\
  &\mathcal{K}^R(x,y)\to \mathcal{K}^R_{\text{M}}(x,y,a,b,c)=\sum_{z=0}^{\min(x,y)}
  \pi_{\text{M}}(x-z,b,c)\pi_\text{H}(z,y,a,b),
  \label{HM1}\\
& \Longrightarrow\mathcal{H}^R(x,y)=\frac1{\sqrt{\pi_\text{M}(x,a+b,c)}}\!
\sum_{z=0}^{\min(x,y)}
  \pi_{\text{M}}(x-z,b,c)\pi_\text{H}(z,y,a,b)
 \sqrt{\pi_\text{M}(y,a+b,c)},
\label{MH1}\\
&\qquad \hat{\phi}_n(x)=d_n\pi_\text{M}(x,a+b,c)
\,{}_2F_1\Bigl(\genfrac{}{}{0pt}{}{-n,\,-x}{a+b}\Bigm|1-c^{-1}\Bigr),\quad d_n^2
=\frac{(a+b)_n\,c^n}{n!}.
\label{M1phn}
\end{align}
\subsubsection{Type (ii) convolution}
\label{sec:Mei2}

By fixing $a$, $b$ with $c\to N(1-c)c^{-1}$ ($\Rightarrow 0<c<1$) and taking
the limit $N\to\infty$ in $\mathcal{K}^R$ of Hahn type (ii) convolution \eqref{HKt2},
one obtains the same Meixner limit $\mathcal{K}^R_{\text{M}}(x,y)$ as in \eqref{HM1}
\begin{equation*}
\mathcal{K}^R(x,y)\to \mathcal{K}^R_{\text{M}}(x,y,a,b,c)
  =\sum_{z=0}^{\min(x,y)}\pi_{\text{M}}(x-z,b,c)\pi_\text{H}(z,y,a,b).
\end{equation*}
\subsubsection{Type (iii) convolution}
\label{sec:Mei3}

This is achieved by fixing $a$ and $c$ with $b\to N(1-b)b^{-1}$
($\Rightarrow 0<b<1$), $N\to\infty$  in $\mathcal{K}^R$ 
of Hahn type (iii) convolution \eqref{HKt3},
\begin{align}
  &\check{P}_n(x,c,a+b)\to\check{P}_{\text{M}\,n}(x,c,b),\n
  &\pi(x,N,c,a+b)\to\pi_{\text{M}}(x,c,b),\quad
  \kappa(n)\to\kappa_{\text{M}}(n)=\frac{(c)_n}{(a+c)_n}
  ={}_2F_1\Bigl(\genfrac{}{}{0pt}{}{-n,\,a}{a+c}\Bigm|1\Bigr),
  \label{Mka3}\\
  &\mathcal{K}^R(x,y)\to \mathcal{K}^R_{\text{M}}(x,y,a,b,c)
  =\sum_{z=\max(x,y)}^{\infty}\pi_\text{H}(x,z,c,a)
  \pi_{\text{M}}(z-y,a,b),
  \label{HM3}\\
 & \Longrightarrow\mathcal{H}^R(x,y)=\frac1{\sqrt{\pi_\text{M}(x,c,b)}}\!
\sum_{z=\max(x,y)}^{\infty}\pi_\text{H}(x,z,c,a)
  \pi_{\text{M}}(z-y,a,b)
 \sqrt{\pi_\text{M}(y,c,b)},
\label{MH3}\\
&\qquad \hat{\phi}_n(x)=d_n\pi_\text{M}(x,c,b)
\,{}_2F_1\Bigl(\genfrac{}{}{0pt}{}{-n,\,-x}{c}\Bigm|1-b^{-1}\Bigr),
\quad d_n^2=\frac{(c)_n\,b^n}{n!}.
\label{M3phn}
\end{align}
\subsection{$q$-Hahn}
\label{sec:qHa}

The $q$-Hahn is defined on a finite integer lattice with two positive parameters
$\bm{\lambda}=(a,b)$ ($0<a<1$, $b<1$) on top of $q$, $0<q<1$  and 
the $q$ dependence of $\pi$ and $\check{P}_n$ is suppressed.
It should be stressed that the polynomial $\check{P}_n(x)$ is a
degree $n$ polynomials in $\eta(x)\eqdef q^{-x}-1$, not in $x$,
\begin{align}
  &\pi(x,N,a,b)=\genfrac{[}{]}{0pt}{}{\,N\,}{x}
  \frac{(a\,;q)_x\,(b\,;q)_{N-x}a^{N-x}}{(ab\,;q)_N},\quad
  \genfrac{[}{]}{0pt}{}{\,N\,}{x}\eqdef\frac{(q\,;q)_N}
  {(q\,;q)_x\,(q\,;q)_{N-x}},
  \label{qHpi}\\[2pt]
  &d_n^2=\genfrac{[}{]}{0pt}{}{\,N\,}{n}
  \frac{(a,abq^{-1}\,;q)_n}{(abq^N,b\,;q)_n\,a^n}\frac{1-abq^{2n-1}}{1-abq^{-1}},
  \label{qHd}\\
  &\check{P}_n(x,a,b)=P_n\bigl(\eta(x),a,b\bigr)
  ={}_3\phi_2\Bigl(\genfrac{}{}{0pt}{}{q^{-n},\,abq^{n-1},\,q^{-x}}
  {a,\,q^{-N}}\Bigm|q\,;q\Bigr).
  \label{qHp}
\end{align}
\subsubsection{Type (i) convolution}
\label{sec:qHa1}
By taking $\bm{\lambda}_1=(a,b)$, $\bm{\lambda}_2=(b,c)$ 
and $\bm{\lambda}_3=(ab,c)$ the matrix
$\mathcal{K}^R(x,y)$ is
\begin{equation}
  \mathcal{K}^R(x,y)=\sum_{z=0}^{\min(x,y)}\pi(x-z,N-z,b,c)\pi(z,y,a,b),  
  \label{qh13K}
\end{equation}
 satisfying
\begin{align}
 & \sum_{y\in\cX} \mathcal{K}^R(x,y)\pi(y,N,ab,c)\check{P}_n(y,ab,c)
  =\kappa(n)\pi(x,N,ab,c)\check{P}_n(x,ab,c),\n
&\qquad    \kappa(n)=\frac{b^n(a\,;q)_n(c\,;q)_n}{(ab\,;q)_n(bc\,;q)_n}
={}_3\phi_2\Bigl(
  \genfrac{}{}{0pt}{}{q^{-n},\,abcq^{n-1},\,b}{ab,\,bc}\Bigm|q\,;q\Bigr),
\label{1qHphi}\\
&\Longrightarrow\mathcal{H}^R(x,y)=\frac1{\sqrt{\pi(x,N,ab,c)}}\!\!\!\!
\sum_{z=0}^{\min(x,y)}\!\!\!\pi(x-z,N-z,b,c)\pi(z,y,a,b)\sqrt{\pi(y,N,ab,c)},
\label{qHH1}\\[2pt]
&\qquad \qquad \hat{\phi}_n(x)=d_n\pi(x,N,ab,c)\,
{}_3\phi_2\Bigl(\genfrac{}{}{0pt}{}{q^{-n},\,abcq^{n-1},\,q^{-x}}
  {ab,\,q^{-N}}\Bigm|q\,;q\Bigr),
\label{qH1ph}\\
&\hspace{2cm}
d_n^2=\genfrac{[}{]}{0pt}{}{\,N\,}{n}
  \frac{(ab,abcq^{-1}\,;q)_n}{(abcq^N,c\,;q)_n\,(ab)^n}\frac{1-abcq^{2n-1}}{1-abcq^{-1}}.
\end{align}

The type (ii) convolution does not exist for $q$-Hahn.
\subsubsection{Type (iii) convolution}
\label{sec:qHa3}
By taking $\bm{\lambda}_1=(a,b)$,  $\bm{\lambda}_2=(c,a)$ 
and  $\bm{\lambda}_3=(c,ab)$ the matrix
$\mathcal{K}^R(x,y)$ is
\begin{equation} 
\mathcal{K}^R(x,y)=\!\!\!\sum_{z=\max(x,y)}^N\!\!\!\pi(x,z,c,a)\pi(z-y,N-y,a,b),
  \label{qh33K}
\end{equation}
 satisfying
\begin{align}
 & \sum_{y\in\cX} \mathcal{K}^R(x,y)\pi(y,N,c,ab)\check{P}_n(y,c,ab)
  =\kappa(n)\pi(x,N,c,ab)\check{P}_n(x,c,ab),\n
&\qquad    \kappa(n)=\frac{a^n(b\,;q)_n(c\,;q)_n}{(ab\,;q)_n(ac\,;q)_n}
={}_3\phi_2\Bigl(\genfrac{}{}{0pt}{}{q^{-n},\,abcq^{n-1},\,a}{ac,\,ab}
  \Bigm|q\,;q\Bigr),
  \label{3qHphi}\\
&\Longrightarrow\mathcal{H}^R(x,y)=\frac1{\sqrt{\pi(x,N,c,ab)}}\!
\sum_{z=\max(x,y)}^N\!\!\!\!\!\!\pi(x,z,c,a)\pi(z-y,N-y,a,b)\sqrt{\pi(y,N,c,ab)},
\label{qHH2}\\[-2pt]
&\qquad \qquad \hat{\phi}_n(x)=d_n\pi(x,N,c,ab)\,
{}_3\phi_2\Bigl(\genfrac{}{}{0pt}{}{q^{-n},\,abcq^{n-1},\,q^{-x}}
  {c,\,q^{-N}}\Bigm|q\,;q\Bigr),
\label{qH2ph}\\
&\hspace{2cm}
d_n^2=\genfrac{[}{]}{0pt}{}{\,N\,}{n}
  \frac{(c,abcq^{-1}\,;q)_n}{(abcq^N,ab\,;q)_n\,c^n}\frac{1-abcq^{2n-1}}{1-abcq^{-1}}.
\end{align}



\end{document}